# Superconductivity in $Mg_{10}Ir_{19}B_{16}$

T. Klimczuk[1,2,*], Q. Xu[3], E. Morosan[1], H. W. Zandbergen[1,3]

and R. J. Cava [1]

[1]Department of Chemistry, Princeton University, Princeton NJ 08544,

[2]Faculty of Applied Physics and Mathematics, Gdansk University of Technology, Narutowicza 11/12, 80-952 Gdansk, Poland,

[3]National Centre for HREM, Kavli Institute of Nanoscience, Delft University of Technology, 2628 CJ Lorentzweg 1, Delft, The Netherlands

## Abstract

$Mg_{10}Ir_{19}B_{16}$, a previously unreported compound in the Mg-Ir-B chemical system, is found to be superconducting at temperatures near 5 K. The fact that the compound exhibits a range of superconducting temperatures between 4 and 5 K suggests that a range of stoichiometries is allowed, though no structural evidence for this is observed. The compound has a large, noncentrosymmetric, body centered cubic unit cell with a = 10.568 Å, displaying a structure type for which no previous superconductors have been reported.

## Introduction

In the past five years, a variety of new superconducting materials have been discovered. Some of these materials are under continuing study because they exhibit unusual properties or structures, e.g. $Na_{0.3}CoO_2{\cdot}1.3H_2O$, which displays superconductivity in a two-dimensional $CoO_2$ lattice,[1] $PuCoGa_5$, the first Pu-based superconductor,[2] $CePt_3Si$, a noncentrosymmetric superconductor,[3] and the pyrochlore superconductors $Cd_2Re_2O_7$,[4] and $MOs_2O_6$ (M=K, Rb).[5,6] Also of interest have been superconducting materials based on light elements such as $MgB_2$,[7] $MgCNi_3$,[8] $C_6Yb$ and $C_6Ca$,[9] and the quaternary intermetallics $LnNi_2B_2C$.[10] In conventional electron-phonon coupled superconductivity, compounds based on light elements are expected to be favorable due to the correspondingly high phonon frequencies they display, suggesting a viable avenue in the search for new superconducting materials.

Here we report the discovery of a new light element based ternary intermetallic superconductor $Mg_{10}Ir_{19}B_{16}$. The compound, synthesized by conventional methods, is superconducting at a critical temperature of about 5 K. A small range of superconducting transition temperatures is observed, indicating that the compound occurs over a range of compositions. This material has a body centered cubic crystal structure with a large unit cell, $a$ = 10.568 Å, and, consequently, a complex formula. There are no previously known superconductors, or intermetallic compounds, with an analogous crystal structure. This paper reports the synthesis and basic superconducting properties of the phase. The details of the crystal structure determination, which has been performed by electron diffraction and imaging methods, will be described elsewhere.[11] The space group of the compound, I-43m, indicates that $Mg_{10}Ir_{19}B_{16}$ is one of the rare examples of a noncentrosymmetric superconducting material.

**Experimental**

During a search in the Mg-Ir-B chemical system, superconductivity was observed to occur in a region of the diagram near the molar ratio 1:2:2. A series of 0.2 g samples was made to optimize the composition of the superconducting phase. Materials were synthesized by standard solid state reaction of pure elements: bright Mg flakes (99% Aldrich Chemical), fine Ir powder (99.9% Alfa Aesar) and amorphous B powder (99.5%, Cerac). After initial mixing, the starting materials were pressed into pellets, wrapped in Ta foil, placed in an $Al_2O_3$ boat, and fired in a quartz-tube furnace under a 95% Ar / 5% $H_2$ atmosphere. The first furnace treatment was for 30 minutes at 600 °C, followed by 1 h at 900 °C. After cooling, the samples were reground with an additional 20% of Mg, pressed into pellets, and placed back in the furnace for 1 hour at 900 °C. The sample purity was checked by powder X-ray diffraction using Cu Kα radiation on a diffractometer equipped with a diffracted beam monochromater. The unit cell parameter was refined from the X-ray data using the profile refinement program TOPAS version 2.1 (Bruker AXS). Zero field cooling dc ($H_{dc}$ = 20 Oe) and ac ($H_{dc}$ = 5 Oe, $H_{ac}$ = 3 Oe, $f$ = 10 kHz) magnetizations were measured in the range of 2 – 6 K (PPMS-Quantum Design). The field dependence of the magnetization was measured at 2 K for one sample. Resistivity measurements were made using a standard ac four-probe technique on pellets cut into dimensions of approximately 2.2 × 0.9 × 0.9 mm.

## Results

The X-ray diffraction patterns of four Mg-Ir-B samples in the vicinity of the optimal composition are presented in Figure 1. The patterns are dominated by a very well crystallized, high symmetry compound, and are very nearly single phase. The vertical bars at the bottom of Figure 1 show the Bragg peak positions for a body centered cubic structure with the refined crystallographic cell parameter $a$ = 10.56782(4) Å. No compound with this lattice type and unit cell dimension has been previously reported in the Mg-Ir-B chemical system, or any related chemical systems.[12] By using the I-43m space group and the unit cell $a$ = 10.56782(4) Å, almost all the peaks in the XRD patterns near the optimal superconducting composition can be indexed. Small amounts of impurities are present, however, in all four samples. Figure 1 shows that the sample with the nominal composition $Mg_{12}Ir_{19}B_{19}$ is very nearly single phase, with less than 5% detectable impurity phase by XRD. Due to the very large crystallographic unit cell, complex formulas are possible for this compound, and the exact composition of the phase can only be determined by detailed structural analysis. These detailed chemical analysis and crystallographic studies, using Electron Probe Microanalysis (EPMA), high resolution electron microscopy (HREM), and quantitative electron diffraction, reported elsewhere,[11] allowed us to determine the space group of the superconducting compound to be I-43m (#217) and the chemical formula to be $Mg_{10}Ir_{19}B_{16}$. This formula is very close to the optimal nominal composition of $Mg_{12}Ir_{19}B_{19}$. The higher amount of Mg required to make the highest purity sample is a result of volatilization of Mg during the synthesis, and is analogous to what was observed in the synthesis of $MgCNi_3$.[8] Similarly, between 15 and 20% excess B is needed to make the purest phase, significantly less than the amount of excess C needed to make optimal $MgCNi_3$ by the same synthetic method. The excess B occurs in elemental form in the sample, and was observed in the HREM investigation.

Figure 2 presents the temperature and magnetic-field dependence of the electrical resistivity of the sample with nominal composition $Mg_{12}Ir_{19}B_{19}$. The main panel shows the temperature range near the superconducting transition. Under zero applied magnetic field, the midpoint of the resistive transition is at 4.4 K, and the 10 – 90 % transition width is 0.4 K. When an external magnetic field is applied, the critical temperature decreases to 3.35 K and 2.4 K at 1 T and 2 T respectively. More measurements under higher magnetic field and at lower temperature would be necessary to make a good estimate of $H_{C2}(0)$, but our results suggest, using the Werthamer-Helfand-Hohenberg (WHH) formula $H_{C2}(0) \cong 0.7T_C(dH_{C2}/dT)\big|_{T_C}$ ,[13] that $H_{C2}(0)$ is low, not exceeding 3 T. The inset of Fig. 2 presents the temperature dependent resistivity between 2 K and 300 K. The normal-state resistivity is rather high, about 3 mΩcm at T = 300 K, about 35 times higher than what is observed for polycrystalline $MgCNi_3$.[8] The shape of ρ(T) and the high value of the resistivity at 300 K suggest that $Mg_{10}Ir_{19}B_{16}$ should be classified as a poor metal. The measurements were performed on a polycrystalline sample, however, with a small amount of elemental boron present in the grain boundaries, and so detailed information about the transport properties of the new superconductor would better be obtained on single crystals.

To characterize the superconducting transition of $Mg_{10}Ir_{19}B_{16}$, zero-field cooling DC magnetization was measured between the temperatures 2 K - 6 K (Figure 3), in an applied field of 20 Oe. The onset of the susceptibility transition is observed at 5 K, a few tenths of a Kelvin higher than the onset of the resistive transition. The superconducting volume fraction from this measurement is estimated to be near 100%. With no detectable impurity phases present beyond the 5% level, this estimated volume fraction and the fact that the diamagnetic signal size is optimized at this composition indicate that the new superconducting compound is the body centered cubic material with composition $Mg_{10}Ir_{19}B_{16}$. The field dependent magnetization curve, M(H), at 2 K is shown in the inset of Figure 3. $H_{C1}$ at 2 K, estimated as the field where the lowest field data deviate from a straight line, is approximately 35 Oe. This low $H_{C1}$ accounts for much of the rounding seen in the measurement of the transition in the applied field of 20 Oe shown in the main panel. With an $H_{C1}$ and $H_{C2}$ of approximately 35 Oe and 30000 Oe, at 2 K, respectively, $Mg_{10}Ir_{19}B_{16}$ is a type 2 superconductor with an estimated $\kappa \approx H_{C2}(2)/H_{C1}(2)$, of 800.

In order to measure the transition temperature more precisely, and its variation with composition, zero field AC susceptibility measurements were performed ($H_{DC}$ = 5 Oe, $H_{AC}$ = 3 Oe, f = 10 kHz). Figure 4 shows the real part of the magnetization, $M'_{AC}$, for the four samples, with nominal compositions $Mg_{10}Ir_{19}B_{18}$, $Mg_{11}Ir_{19}B_{18}$, $Mg_{11}Ir_{19}B_{20}$ and $Mg_{12}Ir_{19}B_{19}$, whose powder X-ray patterns are shown in Figure 1. The data in Figure 4 show that the highest $T_C$ is observed for the sample of nominal composition $Mg_{12}Ir_{19}B_{19}$ - the most chemically pure sample according to the XRD measurements. Almost the same $T_{C\ onset}$, 5 K, is observed for the sample of nominal composition $Mg_{10}Ir_{19}B_{18}$, however the transition is not as sharp as is seen for all other samples, suggesting that the material is chemically inhomogeneous. For $Mg_{11}Ir_{19}B_{20}$ and $Mg_{11}Ir_{19}B_{18}$ the observed diamagnetism is similar, and the transitions are again sharp, but $T_C$ is lower than that observed for the optimal composition: to 4.4 K and 4 K for $Mg_{11}Ir_{19}B_{20}$ and $Mg_{11}Ir_{19}B_{18}$, respectively. The nature of the chemical variability of the superconducting phase is not currently understood, however, the fact that very little variation in lattice parameter is observed suggests both that the chemical variability is small and that it likely involves the boron, which is much smaller than the metallic constituents Mg and Ir, and is expected to change the unit cell size only slightly if variable.

**Conclusion**

By exploring the Mg-Ir-B phase diagram in a search for light-element based intermetallic superconductors, a new ternary superconducting compound was discovered. The new intermetallic compound, $Mg_{10}Ir_{19}B_{16}$, has a cubic structure with a large, body centered unit cell of $a$ = 10.568 Å. The synthesis of this compound is similar to that used in the case of $MgCNi_3$ and requires of excess of both Mg and B in the starting material. In its normal state, $Mg_{10}Ir_{19}B_{16}$ is a poor metal with high electrical resistivity at room temperature. The critical temperature, $T_C$ about 5 K, was determined by both electrical resistivity and magnetization measurements. Finally $Mg_{10}Ir_{19}B_{16}$ is a rare example of noncentrosymmetric superconducting material, and has a new intermetallic structure type. The characterization of related materials with the same structure type would be of significant interest.

**Acknowledgements**

This work was supported by the US Department of Energy, grant DE-FG02-98-ER45706. The work at Delft was supported by The Nederlandse Stichting voor Fundamenteel Onderzoek der Materie (FOM). Tomasz Klimczuk would like to thank The Foundation for Polish Science for support.

* Present address: Los Alamos National Laboratory, Los Alamos, New Mexico 87545, USA

**Figure Captions**

**Figure 1.** Powder X-ray diffraction patterns ($CuK_{\alpha}$ radiation) for $Mg_{10}Ir_{19}B_{18}$ (A), $Mg_{11}Ir_{19}B_{18}$ (B), $Mg_{11}Ir_{19}B_{20}$ (C) and $Mg_{12}Ir_{19}B_{19}$ (D). The latter composition is the most nearly phase pure. Vertical bars at the bottom represent the Bragg peak positions for a body centered cubic cell with refined cell parameter $a$ = 10.56782 Å. Miller indices for each peak are shown, and impurity phase peak positions are marked with arrows.

**Figure 2.** Magnetic-field dependence of the electrical resistivity for $Mg_{10}Ir_{19}B_{16}$ (nominal composition $Mg_{12}Ir_{19}B_{19}$) – detail near the superconducting transition. The inset shows the temperature dependent resistivity between 2 K and 300 K

**Figure 3.** DC magnetization characterization of the superconducting transition in $Mg_{10}Ir_{19}B_{16}$ (nominal composition $Mg_{12}Ir_{19}B_{19}$). The inset shows magnetization vs. applied magnetic field dependence at 2 K.

**Figure 4.** AC magnetization characterization of the superconducting transition for samples with nominal compositions $Mg_{10}Ir_{19}B_{18}$, $Mg_{11}Ir_{19}B_{18}$, $Mg_{11}Ir_{19}B_{20}$ and $Mg_{12}Ir_{19}B_{19}$.

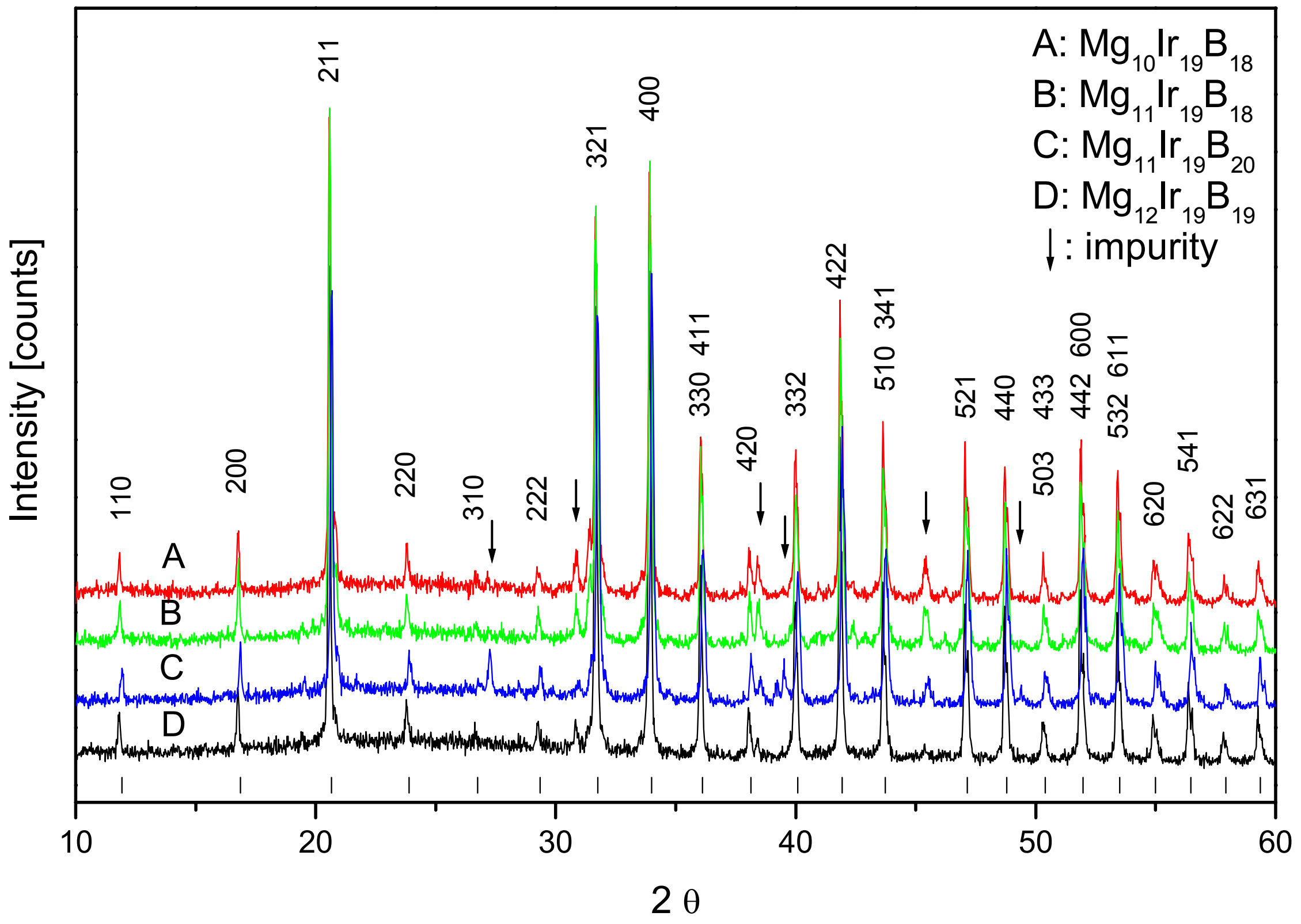

**Fig. 1**

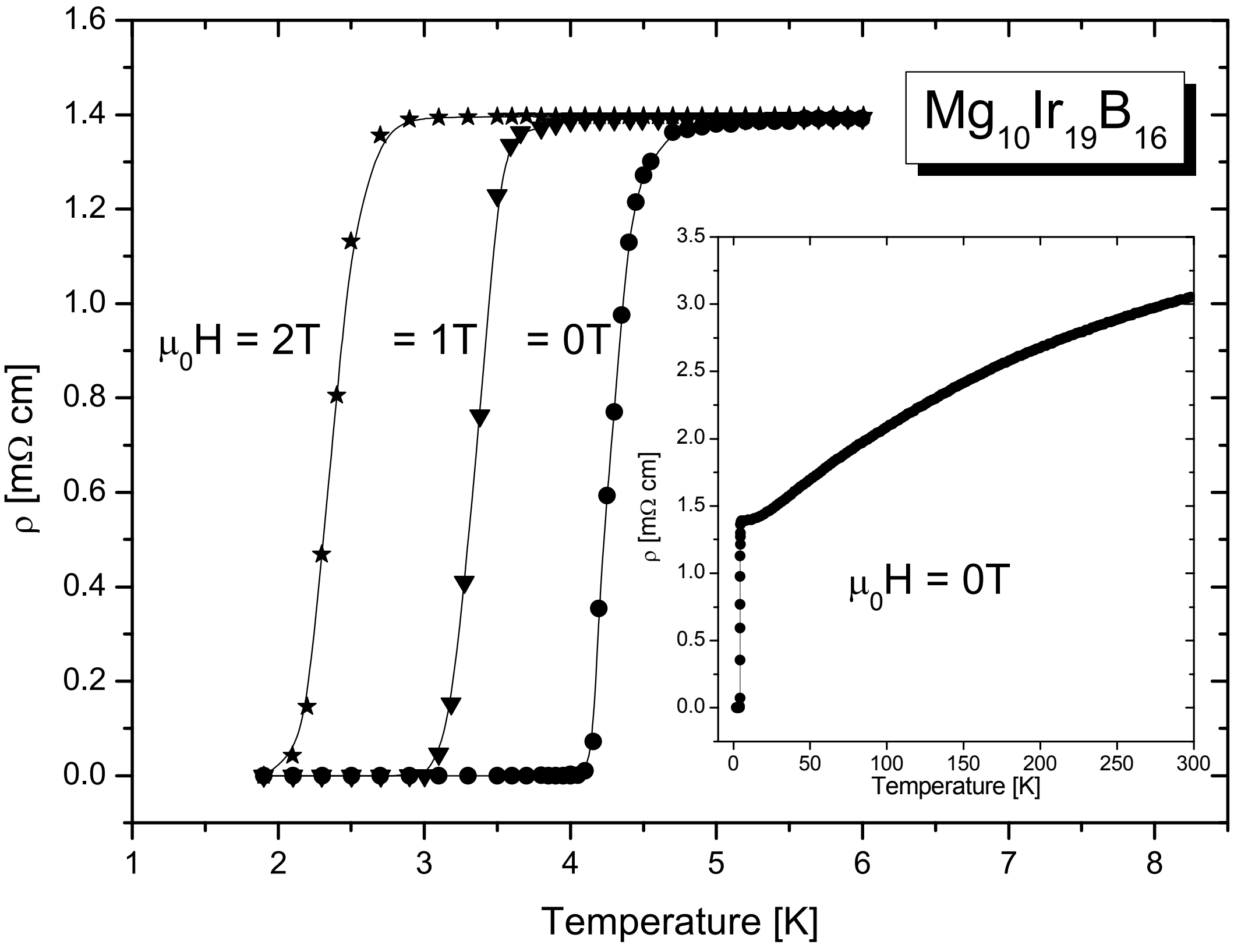

**Fig. 2**

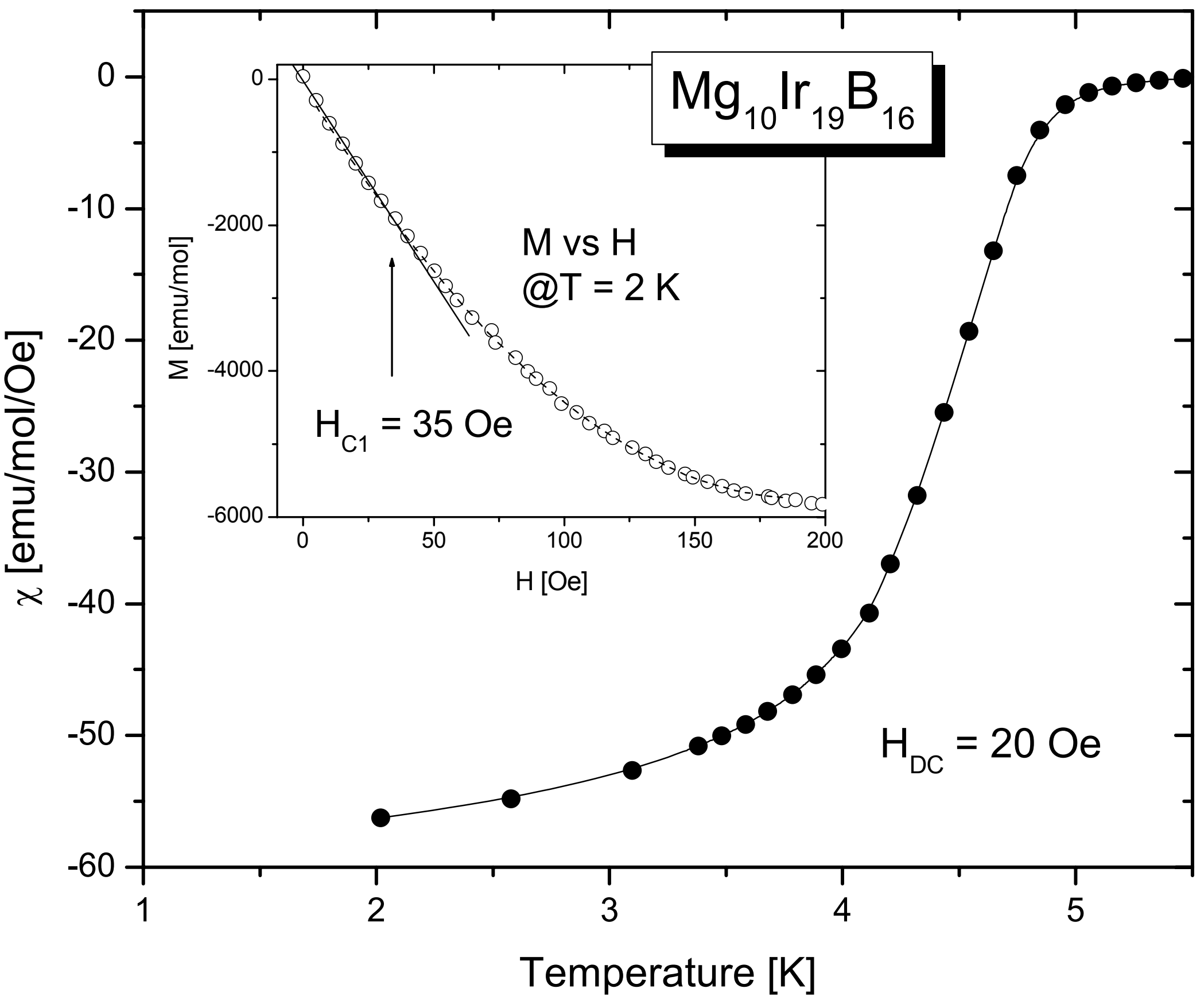
Mg10Ir19B16
M vs H
@T = 2 K
HC1 = 35 Oe
M [emu/mol]
H [Oe]
HDC = 20 Oe
χ [emu/mol/Oe]
Temperature [K]

**Fig. 3**

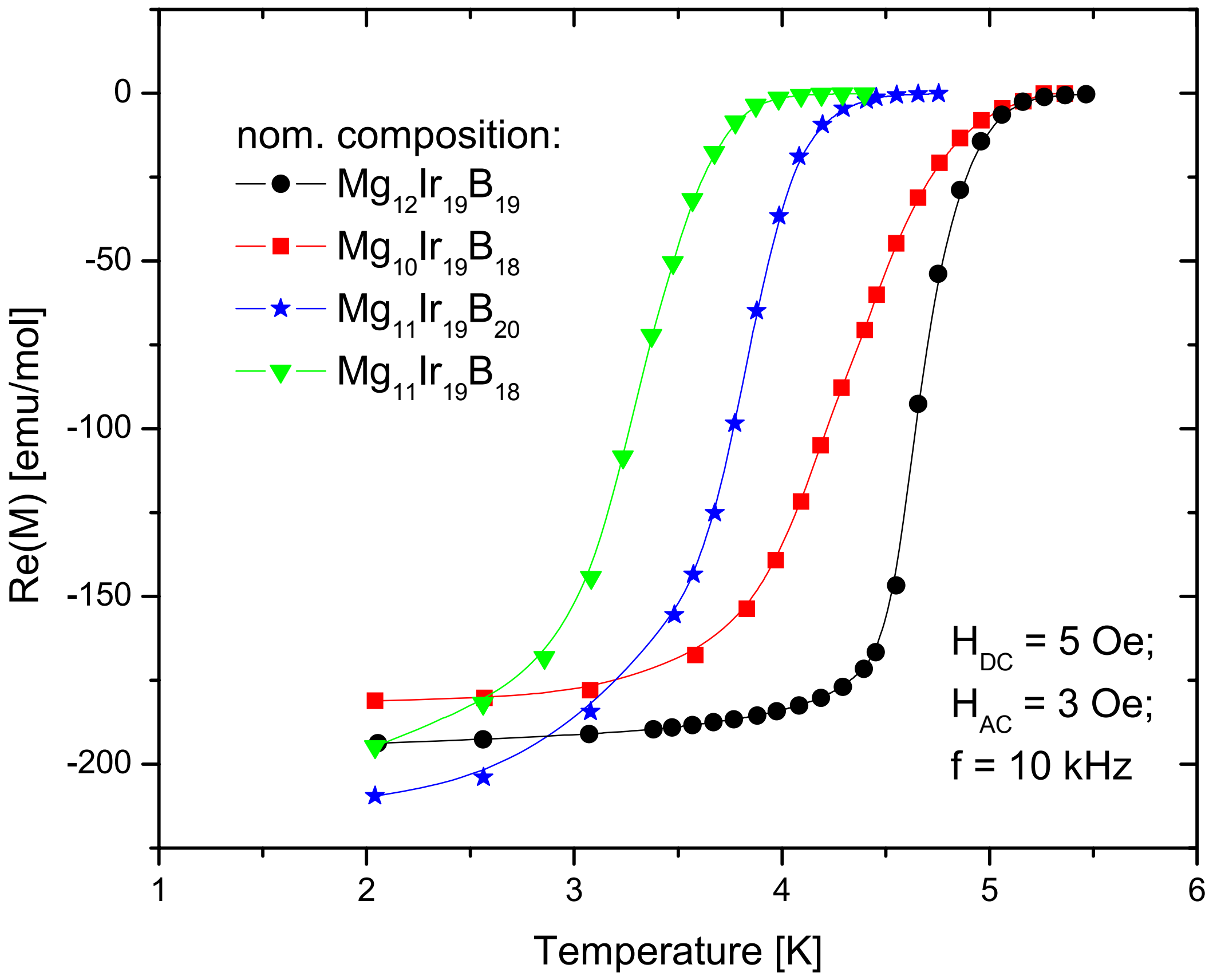

0
-50
-100
-150
-200
1
2
3
4
5
6
Re(M) [emu/mol]
Temperature [K]
nom. composition:
Mg12Ir19B19
Mg10Ir19B18
Mg11Ir19B20
Mg11Ir19B18
HDC = 5 Oe;
HAC = 3 Oe;
f = 10 kHz

**Fig. 4**